\documentclass[aps,prl,showpacs,twocolumn]{revtex4}

\usepackage{amssymb}
\usepackage{epsfig}
\usepackage{graphicx}
\usepackage{amsmath}
\usepackage{array,color}

\begin{document}

\title{Quantized Quasi-Two Dimensional Bose-Einstein Condensates with Spatially Modulated Nonlinearity}
\author{Deng-Shan Wang$^1$, Xing-Hua Hu$^1$, Jiangping Hu$^2$ and W. M. Liu$^1$}
\address{$^1$Beijing National Laboratory for Condensed Matter Physics,
Institute of Physics, Chinese Academy of Sciences, Beijing $100190$,
P.R. China}
\address{$^2$Department of Physics, Purdue University, West Lafayette, Indiana $47907$, U.S.A.}
\date{\today}

\begin{abstract}
We investigate the localized nonlinear matter waves of the quasi-two
dimensional Bose-Einstein condensates with spatially modulated
nonlinearity in harmonic potential. It is shown that the whole
Bose-Einstein condensates, similar to the linear harmonic
oscillator, can have an arbitrary number of localized nonlinear
matter waves with discrete energies, which are mathematically exact
orthogonal solutions of the Gross-Pitaevskii equation. Their novel
properties are determined by the principle quantum number $n$ and
secondary quantum number $l$: the parity of the matter wave
functions and the corresponding energy levels depend only on $n$,
and the numbers of density packets for each quantum state depend on
both $n$ and $l$ which describe the topological properties of the
atom packets. We also give an experimental protocol to observe these
novel phenomena in future experiments.

\end{abstract}

\pacs{03.75.Hh, 05.45.Yv,  67.85.Bc}

\maketitle \emph {Introduction}.---Since the remarkable experimental
realization \cite{Anderson,Hulet1,Davis} of Bose-Einstein
condensations (BEC), there has been an explosion of the experimental
and theoretical activity devoted to the physics of dilute ultracold
bosonic gases.
 It is known that the properties of BEC
including their shape, collective nonlinear excitations are
determined by the sign and magnitude of the $s$-wave scattering
length. A prominent way to adjust scattering length is to tune an
external magnetic field in the vicinity of a Feshbach resonance
\cite{Inouye1}. Alternatively, one can use a Feshbach resonance
induced by optical or electric field \cite{Theis}. Since all
quantities of interest in the BEC crucially depend on scattering
length, a tunable interaction suggests very interesting studies of
the many-body behavior of condensate systems.
\par
 In the past
years, techniques for adjusting the scattering length globally have
been crucial to many experimental achievements
\cite{Herbig,Bartenstein}. More recently, condensates with a
spatially modulated nonlinearity by manipulating scattering length
locally have been proposed
\cite{Rodas-Verde,Sakaguchi,Konotop,Qian,Belmonte-Beitia12}.  This
is experimentally feasible due to the flexible and precise control
of the scattering length with tunable interactions. The spatial
dependence of scattering length can be implemented by a spatially
inhomogeneous external magnetic field in the vicinity of a Feshbach
resonance \cite{Xiong}.
\par
However, so far, the studies of BEC with spatially modulated
nonlinearity are limited in the quasi-one dimensional cases
\cite{Rodas-Verde,Sakaguchi,Konotop,Qian,Belmonte-Beitia12}.
Moreover, in the study of nonlinear problems no one discusses their
quantum properties which are common in linear systems such as the
linear harmonic oscillator. In this Letter, we extend the similarity
transformation \cite{Belmonte-Beitia12} to the quasi-two dimensional
(quasi-2D) BEC
 with spatially modulated nonlinearity in harmonic potential, and
find a family of stable localized nonlinear matter wave solutions.
Similar to the linear harmonic oscillator, we discover that the
whole BEC can be quantized which is unexpected before. Their quantum
and topological properties can be simply described by two quantum
numbers. We also formulate an experimental procedure for the
realization of these novel phenomena in $^{7}$Li condensate
\cite{Hulet1,Hulet2}. This opens the door to the investigation of
new matter waves in the high dimensional BEC with spatially
modulated nonlinearities.
\par
{\em Model and exact localized solutions}.---The system considered
here is a BEC confined
 in a harmonic trap $V(\textbf{r})=m(\omega_\perp^2 r^2+\omega_z^2z^2)/2$,
 where $m$ is atomic mass, $r^2=x^2+y^2$, and $\omega_{\perp}, \omega_{z}$ are the confinement
frequencies in the radial and axial directions, respectively. In the
mean-field theory, the BEC system at low temperature is described by
the Gross-Pitaevskii (GP) equation in three dimensions. If the trap
is pancake-shaped, i.e. $\omega_z\gg\omega_\perp,$ it is reasonable
to reduce the GP equation for the condensate wave function to a
quasi-2D equation \cite{Kivshar,Ueda1,Garcia-Ripoll}
\begin{equation}\label{GP}
i\psi_t=-\frac{1}{2}(\psi_{xx}+\psi_{yy})+\frac{1}{2}\omega^2(x^2+y^2)\psi+g(x,y)|\psi|^2\psi,
\end{equation}
where $\omega=\omega_\perp/\omega_z$, the length, time and wave
function $\psi$ are measured in units of
$a_h=\sqrt{\hbar/m\omega_z}, \omega_z^{-1}, a_h^{-1}$ and
$g(x,y)=4\pi a_s(x,y)$ represents the strength of interatomic
interaction characterized by the $s$-wave scattering length
$a_s(x,y)$, which can be spatially inhomogeneous by magnetically
tuning the Feshbach resonances
\cite{Inouye1,Rodas-Verde,Sakaguchi,Konotop,Qian,Belmonte-Beitia12,Xiong}.
\begin{figure}[tbp]
\includegraphics[width=9.0cm]{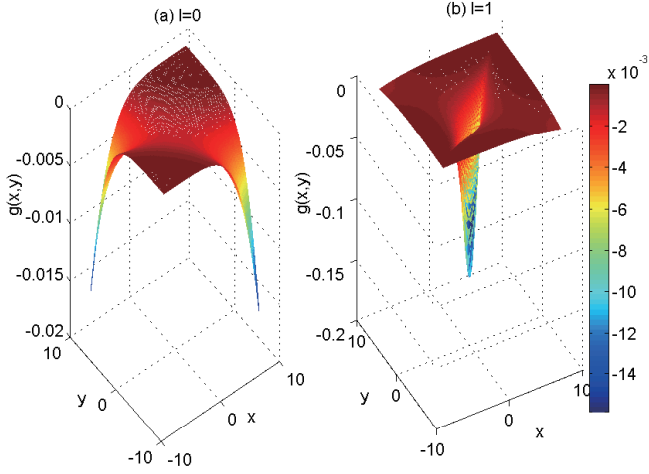}
\caption{\small (color online). The interaction parameter $g(x,y)$
for two secondary quantum numbers: $(a)~l=0$ and $(b)~l=1$ with
$\omega=0.02, \nu=0.1$. It is seen that $g(x,y)$ is a smooth
function when $l=0$ and develops singularity when $l$ gets large. }
\end{figure}

Now we consider the spatially localized stationary solution
$\psi(x,y,t)=\phi(x,y)e^{-i\mu t}$ of Eq. (\ref{GP}) with
$\phi(x,y)$ being a real function for $\lim_{|x|,|y|\rightarrow
\infty}\phi(x,y)=0.$ This maps Eq. (\ref{GP}) onto a stationary
nonlinear Schr\"{o}dinger equation
$\frac{1}{2}\phi_{xx}+\frac{1}{2}\phi_{yy}-\frac{1}{2}\omega^2(x^2+y^2)\phi-g(x,y)\phi^3+\mu
\phi=0$ \cite{Pethick}. Here $\mu$ is the real chemical potential.
Solving this stationary equation by similarity transformation
\cite{Belmonte-Beitia12}, we obtain a families of exact localized
nonlinear wave solutions for Eq. (\ref{GP}) as
\begin{equation}
\psi_n= {\frac {(n+1)K(k)\eta }{ \sqrt {\nu}}}\,{\rm cn}
(\theta,k)e^{-i\mu t},n=0,2,4,\cdots
\end{equation}
\begin{equation}
 \psi_n={\frac {
(n+1)K(k)\eta }{\sqrt {2\nu}}}\,{\rm sd} (\theta,k )e^{-i\mu
t},n=1,3,5,\cdots
\end{equation}
where $k=\sqrt {2}/{2}$ is the modulus of elliptic function, $\nu$
is a positive real constant,
$K(k)=\int_{0}^{\frac{\pi}{2}}[1-k^2\sin^2 \varsigma]d\varsigma$ is
elliptic integral of the first kind,  ${\rm sd}={\rm sn}/{\rm dn}$
with ${\rm sn},{\rm cn}$ and ${\rm dn}$ being Jacobi elliptic
functions, $\theta,\eta$ and $g$ are determined by
$$
\theta=(n+1)K(k){\rm erf}[\sqrt {2\omega} \left( x+y \right)/2],
$$
\begin{equation}
\eta={e^{\omega\,xy}}{\rm KummerU} [ -\mu/(2\omega),1/2,\omega\,
\left( x-y\right) ^{2}/2],
\end{equation}
$$g(x,y)=-2\omega\,\nu/(\pi \eta ^{2})
 {e^{-\omega\, \left( x+y
 \right) ^{2}}},
$$
here ${\rm erf}(x)=\frac{2}{\sqrt{\pi}}\int_0^{x}e^{-\tau^2}d\tau$
is error function, and ${\rm KummerU}(a,c,s)$ \cite{Abramowitz} is
Kummer function of the second kind which is a solution of ordinary
differential equation
$s\Lambda^{''}(s)+(c-s)\Lambda^{'}(s)-a\Lambda(s)=0.$ It is easy to
see that when $|x|,|y|\rightarrow \infty$ we have $\psi_n\rightarrow
0$ for solutions $\psi_n$ in Eqs. (2)-(3) with Eq. (4), thus they
are localized bound state solutions.
\par
In the above construction, it is observed that the number of zero
points of function $\eta$ in Eq. (4) is equal to that of function
${\rm KummerU} [ -\mu/(2\omega),1/2,\omega\, \left( x-y\right)
^{2}/2],$ which strongly depends on $\omega$ and the ratio
$\mu/\omega.$ We assume the number of zero points in $\eta$ along
line $y=-x$ is $l.$ In the following, we will see that integer $n$
is associated with the energy levels of the atoms and integers $n,l$
determine the topological properties of atom packets, so $n$ and $l$
are named the principal quantum number and secondary quantum number
in quantum mechanics. In addition, the three free parameters
$\omega$, $\mu$ and $\nu$ are positive, so the dimensionless
interaction function $g(x,y)$ is negative, which indicates an
attractive interaction between atoms. There are known atomic gases
with attractive interactions realized by modulating magnetic
\cite{Inouye1} technique, for examples, the $^{85}$Rb \cite{Cornish}
and $^{7}$Li atoms \cite{Hulet1,Hulet2}.
\begin{figure}[tbp]
\includegraphics[width=9.4cm]{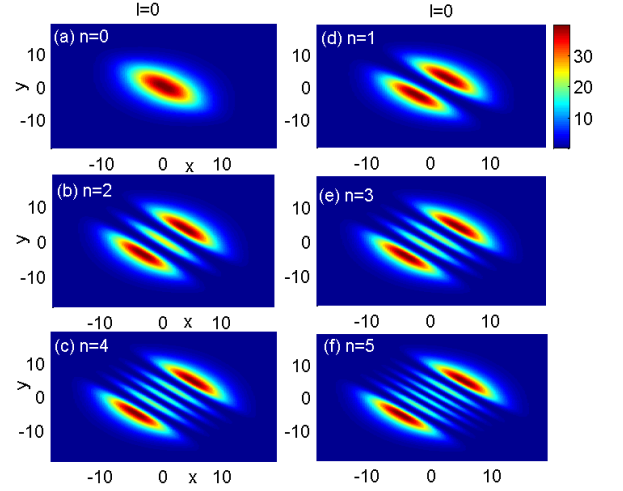}
\caption{\small (color online). The density distributions of the
quasi-2D BEC with spatially modulated nonlinearities in harmonic
potential, for different principle quantum numbers $n$ in Eqs.
(2)-(3) with Eq. (4), where the secondary quantum number $l=0,$ the
parameters $\omega,\nu$ are $0.02$ and $0.1,$ respectively. The unit
of space length $x,y$ is $1.69~\mu m.$ Figs. 2(a)-2(c) show the
density profiles of the even parity wave function (2) for $n=0,2$
and $4,$ respectively. Figs. 2(d)-2(f) demonstrate the density
profiles of the odd parity wave function (3) for $n=1,3$ and $5,$
respectively. }
\end{figure}
\par
To translate our results into units relevant to the experiments
\cite{Hulet1,Hulet2}, we take the $^{7}$Li condensate containing
$10^3\sim 10^5$ atoms in a pancake-shaped trap with radial frequency
$\omega_\perp=2\pi \times 10$ Hz and axial frequency $\omega_z=2\pi
\times 500$ Hz \cite{Rychtarik}. In this case, the ratio of trap
frequency $\omega$ in Eq. (1) is $0.02$ which is determined by
$\omega_\perp/\omega_z.$ The unit of length is $1.69~\mu m$, the
unit of time is $0.32~ ms$ and the unit of chemical potential is
$nK$. The spatially inhomogeneous interaction parameter $g(x,y)$ is
independent of principal quantum number $n$ but is strongly related
to the secondary quantum number $l$. In the Fig. 1, we show that for
$\omega=0.02, \nu=0.1,$ function $g(x,y)$ is smooth in space when
$l=0$ and develops singularity when the $l$ gets large.
\par
{\em Quantized quasi-$2$D BEC.}---In order to investigate the
quantum and topological properties of the localized nonlinear matter
waves in quasi-2D BEC described by Eqs. (2)-(3) with Eq. (4), we
plot their density distributions by manipulating the principal
quantum number $n$ or secondary quantum number $l$.
\begin{figure}[tbp]
\includegraphics[width=10.3cm]{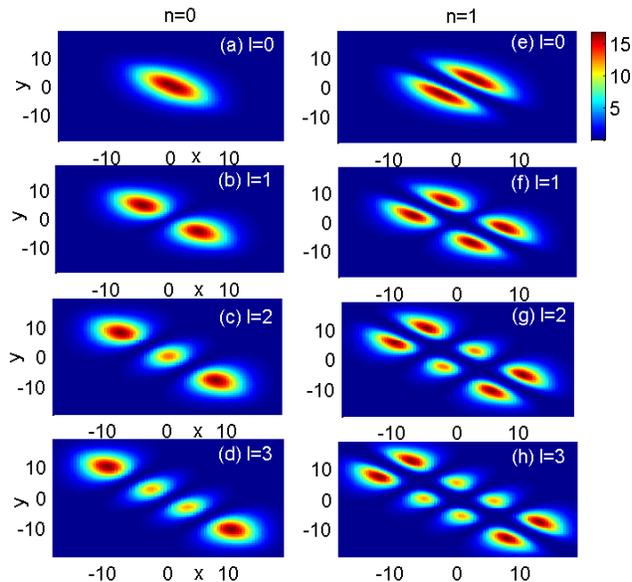}
\caption{\small (color online). The density distributions of the
quasi-2D BEC in harmonic potential for different secondary quantum
number $l$. Figs. 3(a)-3(d) show the density profiles of the even
parity wave function (2) for principle quantum number $n=0,$ and
Figs. 3(e)-3(h) show the density profiles of the odd parity wave
function (3) for $n=1,$ corresponding to $l=0,1,2,3$. The other
parameters are the same as that of Fig. 2. }
\end{figure}
\par
 Firstly, when the secondary quantum number $l$ is fixed, we can
modulate the principal quantum number $n$ to analyze the novel
matter waves in quasi-2D BEC. Fig. 2 shows the density profiles in
quasi-2D BEC with spatially modulated nonlinearities in harmonic
potential for $l=0$. It is easy to see that the matter wave
functions in Eq. (2) satisfy $\psi_n(-x,-y)=\psi_n(x,y),$ so they
are even parity and are invariant under space inversion. Figs.
2(a)-2(c) demonstrate the density profiles of the even parity wave
functions (2) with Eq. (4) for $n=0,2,4,$ which correspond to a low
energy state and two highly excited states. The numbers of atoms
$N_n=\int\int dxdy |\psi_n(x,y,t)|^2$ for the three states are
$N_0=3.76\times 10^3, N_2= 6.84\times 10^4, N_4=2.633\times10^5$,
respectively. The matter wave functions in Eq. (3) satisfy
$\psi_n(-x,-y)=-\psi_n(x,y),$ which denotes that they are odd
parity. Figs. 2(d)-2(f) demonstrate the density profiles of the odd
parity wave functions (3) with Eq. (4) for $n=1,3,5,$ which
correspond to three highly excited states. The numbers of atoms for
the three states are $N_1=4.016\times10^4, N_3=2.493\times 10^5,
N_5=7.28\times10^5,$ respectively. It is observed that when the
secondary quantum number $l=0$, the number of nodes along line $y=x$
for each quantum state is equal to the corresponding principal
quantum number $n$, i.e. the $n$th level quantum state has $n$ nodes
along $y=x$. And the number of density packets increases one by one
along line $y=x$ when the $n$ increases. This is similar to the
quantum properties in the linear harmonic oscillator.
\par
Secondly, when the principal quantum number $n$ is fixed, we can
tune the secondary quantum number $l$ to observe the novel quantum
phenomenon in quasi-2D BEC. In Fig. 3 we demonstrate the density
distributions of quasi-2D BEC in harmonic potential for different
secondary quantum number. Figs. 3(a)-3(d) show the density profiles
of the even parity wave function (2) with Eq. (4) for $n=0,$ and
$l=0,1,2$ and $3,$ respectively. It is seen that the number of nodes
for the density packets along line $y=-x$ is equal to the
corresponding secondary quantum number $l$ which describes the
topological patterns of the atom packets, and the number of density
packets increases one by one when $l$ increases. Figs. 3(e)-3(h)
show the density profiles of the odd parity wave function (3) with
Eq. (4) for $n=1$ and $l=0,1,2,3.$ We see that the number of density
packets increases pair by pair when $l$ increases. The number of
density packets for each quantum state is equal to $(n+1)\times
(l+1),$ and all the density packets are symmetrical with respect to
lines $y=\pm x,$ as shown in Figs. 2-3.

\par
{\em Normalization energy vs chemical potential.} Next we calculate
the normalization energy of each quantum states numerically. The
total energy of the quasi-2D BEC is $E(\psi)=\int\int d xdy[|\nabla
\psi|^2+\frac{1}{2}\omega^2(x^2+y^2)|\psi|^2+\frac{1}{2}g(x,y)|\psi|^4]$.
So the normalized energy is given by
$E(\psi)/N=\mu-\frac{1}{2N}\int\int dxdy g(x,y)|\psi|^4$ with
$N=\int\int d xdy |\psi|^2$. Fig. 4 shows the relations of the
normalization energy $E(\psi)/N$ with chemistry potential for
different principle quantum numbers $n$. It is observed that for the
fixed $n$, the normalization energy is approximatively linear
increase with respect to chemistry potential, i.e.,
$d(E(\psi)/N)/d\mu>0.$ Fig. 4(a) demonstrates that the normalization
energy for the even parity wave function (2) increases when the
principal quantum number $n$ increases. So does the odd parity wave
function (3), as shown in Fig. 4(b). It is shown that the energy
levels of the atoms are only associated with the principle quantum
number $n$. These are similar to energy level distribution of the
energy eigenvalue problem for the linear harmonic oscillator
described by linear Schr\"{o}dinger equation.

\begin{figure}[tbp]
\includegraphics[width=8.5cm]{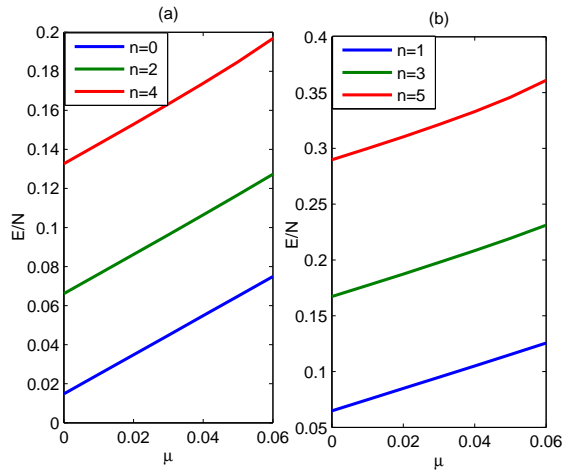}
\caption{\small (color online). The normalization energy $E(\psi)/N$
vs chemical potential $\mu$, with $N=\int \int d xdy |\psi|^2$. (a)
even parity wave function (2) with principal quantum numbers
$n=0,2,4$ and (b) odd parity wave function (3) with $n=1,3,5$. Here
the parameters $\omega=0.02$ and $\nu=0.1.$}
\end{figure}
\par
{\em Stability analysis.}---Stability of exact solutions with
respect to perturbation is very important, because only stable
localized nonlinear matter waves are promising for experimental
observations and physical applications. To study the stability of
our exact solutions (2)-(3) with Eq. (4), we consider a perturbed
solution $\psi(x,y,t)=[\phi_n(x,y)+\Psi(x,y,t)]e^{-i\mu t}$ of Eq.
(1). Here $\phi_n(x,y)$ are the exact solutions of the stationary
nonlinear Schr\"{o}dinger equation
$\frac{1}{2}\phi_{xx}+\frac{1}{2}\phi_{yy}-\frac{1}{2}\omega^2(x^2+y^2)\phi-g(x,y)\phi^3+\mu
\phi=0$. $\Psi(x,y,t)\ll 1$ is a small perturbation to the exact
solutions and $\Psi(x,y,t)=[R(x,y)+I(x,y)]e^{i\lambda t}$ is
decomposed into its real and imaginary parts \cite{Bronski}.
Substituting this perturbed solution to the quasi-2D GP equation (1)
and neglecting the higher-order terms in $(R,I)$, we obtain a
standard eigenvalue problem $ L_{+}R=\lambda I,~L_{-}I=\lambda R,$
where $\lambda$ is eigenvalue, $R,I$ are eigenfunctions with
$L_{+}=-\frac{1}{2}(\partial_x^2+\partial_y^2)+3g(x,y)\phi_n(x,y)^2+\frac{1}{2}\omega^2(x^2+y^2)-\mu$
and
$L_{-}=-\frac{1}{2}(\partial_x^2+\partial_y^2)+g(x,y)\phi_n(x,y)^2+\frac{1}{2}\omega^2(x^2+y^2)-\mu.$
Numerical experiments show that when $\omega=0.02$ and $\mu,\nu $
are arbitrary non-negative constants, only for principle quantum
number $n=0,1,2,3,4,5$ are the eigenvalues $\lambda$ of this
eigenvalue problem real. This suggests that for $\omega=0.02$ the
exact localized nonlinear matter wave solution (2) is linear
stability only for $n=0,2,4$ and solution (3) is linear stability
only for $n=1,3,5,$ see Fig. 5. It is seen that when the frequencies
of pancake-shaped trap is fixed, the stability of the exact
solutions (2)-(3) with Eq. (4) rests only on the principle quantum
number $n.$
\par
 {\em Experimental protocol.}---We now
provide an experimental protocol for creating the quasi-2D localized
nonlinear matter waves. To do so, we take the attractive $^{7}$Li
condensate \cite{Hulet1,Hulet2}, containing about $10^3\sim10^5$
atoms, confined in a pancake-shaped trap with radial frequency
$\omega_\perp=2\pi \times 10$ Hz and axial frequency $\omega_z=2\pi
\times 500$ Hz \cite{Rychtarik}. This trap can be determined by
combination of spectroscopic observations, direct magnetic field
measurement, and the observed spatial cylindrical symmetry of the
trapped atom cloud \cite{Rychtarik}. The next step is to realize the
spatial variation of the scattering length. Near the Feshbach
resonance \cite{Inouye1,Xiong,Ueda1,Khaykovich}, the scattering
length $a_s(B)$ varies dispersively as a function of magnetic field
$B,$ i.e. $a_s(B)=a[1+\Delta/(B_0-B)],$ with $a$ being the
asymptotic value of the scattering length far from the resonance,
$B_0$ being the resonant value of the magnetic field, and $\Delta$
being the width of the resonance. For the magnetic field in $z$
direction with gradient $\alpha$ along $x$-$y$ direction, we have
$\vec{B}=[B_0+\alpha B_1(x,y)]\vec{e_z}$. In this case, the
scattering length is dependent on $x$ and $y$. In real experiments,
the spatially dependent magnetic field may be generated by a
microfabricated ferromagnetic structure integrated on an atom chip
\cite{Vengalattore1,Vengalattore2}, such that interaction in Fig. 1
can be realized. In order to observe the density distributions in
Figs. 2-3 clearly in experiment, the $^{7}$Li atoms should be
evaporatively cooled to low temperatures, say in the range of 10 to
100 $nK$. After the interaction parameter in Fig. 1(a) is realized
by modulating magnetic field properly, the density distributions in
Fig. 2 can be observed for different numbers of atoms by evaporative
cooling, for example, the numbers of atoms in Fig. 2(a)-2(c) are
$3.76\times 10^3, 6.84\times 10^4, 2.633\times10^5$, respectively.
The density distributions in Fig. 3 can also be observed by changing
the scattering lengths through magnetic field for various atom
numbers.

\begin{figure}[tbp]
\includegraphics[width=9.5cm]{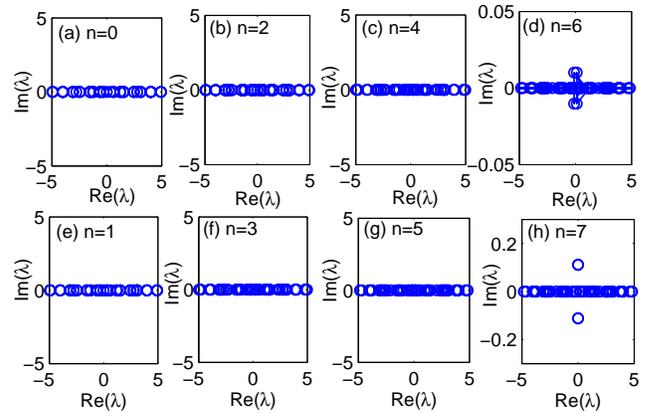}
\caption{\small (color online). Eigenvalue for different principal
quantum number $n$ with parameters $\omega=0.02,\mu=0.001$ and
$\nu=0.1.$ It is shown that only for $n=0,1,2,3,4,5$ are the
localized nonlinear matter wave solutions (2)-(3) with Eq. (4)
linear stability.}
\end{figure}
\par
 {\em Conclusion}.---In summary, we have discovered a new family of stable exact localized
 nonlinear matter wave solutions of the quasi-2D BEC with spatially
modulated nonlinearities in harmonic potential. Similar to the
linear harmonic oscillator, we introduce two classes of quantum
numbers: the principle quantum number $n$ and secondary quantum
number $l$. The matter wave functions have even parity for the even
principle quantum number and odd parity for the odd one, the energy
levels of the atoms are only associated with the principle quantum
number, and the number of density packets for each quantum state is
equal to $(n+1)\times (l+1)$. We also provide an experimental scheme
to observe these novel phenomena in future experiments. Our results
are of particular significance to matter wave management in high
dimensional BEC.
\par
This work was supported by NSFC under Grants No. 10874235, No.
10934010, No. 60978019 and by NKBRSFC under Grants No. 2006CB921400,
No. 2009CB930704 and No. 2010CB922904.

\end{document}